# Real-Time Measurements of Photonic Microchips with Femtometer-Scale Spectral Precision and Ultra-High Sensitivity

Mahdi Mozdoor Dashtabi[1,*], Mohammad Talebi Khoshmehr[1], Hamed Nikbakht[1], Bruno Lopez Rodriguez[2], Naresh Sharma[2], Iman Esmaeil Zadeh[2], and B. Imran Akca[1,**]

[1]LaserLab, Department of Physics and Astronomy, VU University, De Boelelaan 1081, 1081 HV, Amsterdam, The Netherlands
[2]Department of Imaging Physics (ImPhys), Faculty of Applied Sciences, Delft University of Technology, Delft, The Netherlands

m.mozdoor.dashtabi@vu.nl; m.talebi.khoshmehr@vu.nl; h.nikbakht@vu.nl; b.lopezrodriguez@tudelft.nl, N.Sharma-1@tudelft.nl; I.EsmaeilZadeh@tudelft.nl; b.i.avci@vu.nl

*e-mail: m.mozdoor.dashtabi@vu.nl
**e-mail: b.i.avci@vu.nl

**ABSTRACT:** Photonic integrated circuits (PICs) are enabling major breakthroughs in a number of areas, including quantum computing, neuromorphic processors, wearable devices, and more. Nevertheless, existing PIC measurement methods lack the spectral precision, speed, and sensitivity required for refining current applications and exploring new frontiers such as point-of-care or wearable biosensors. Here, we present the "Sweeping Optical Frequency Mixing Method (SOHO)", surpassing traditional PIC measurement methods with real-time operation, 30 dB higher sensitivity, and over 100 times better spectral resolution. Leveraging the frequency mixing process with a sweeping laser and custom control software, SOHO excels in simplicity, eliminating the need for advanced optical components and additional calibration procedures. We showcase its superior performance on ultrahigh-quality factor ($Q$) fiber-loop resonators ($Q = 46 \times 10^6$) as well as microresonators realized on a new optical waveguide platform. An experimental spectral resolution of 19.1 femtometers is demonstrated using an 85-meter-long unbalanced fiber Mach Zehnder Interferometer, constrained by noise resulting from the extended fiber length, while the theoretical resolution is calculated to be 6.2 femtometers, limited by the linewidth of the reference laser. With its excellent performance metrics, SOHO has the potential to become a vital measurement tool in photonics, excelling in high-speed and high-resolution measurements of weak optical signals.

## INTRODUCTION

Photonic integration is revolutionizing the field of optics in the same way that integrated circuits revolutionized the world of electronics in the 1960s. Miniature optical circuits known as photonic integrated circuits (PIC) leverage the principles of photonics to integrate multiple optical components on a single chip. PICs are becoming increasingly common in quantum computing, data centers, lidar and autonomous vehicles, neuromorphic processors, medical diagnostics, optical sensing, and astronomy. In particular, ultra-high quality factor microresonators ($Q \geq 10^6$) have been extensively employed in various emerging applications including optical sensing[1,2,3,4], photonic processors[5,6], ultra-narrow linewidth lasers[7,8], precision spectroscopy[9,10], quantum optics[11,12,13], optical filters[14,15,16] and nonlinear optics[17,18,19,20,21]. Real-time and precise optical measurements of these components are not only crucial for optimizing their performance in existing applications but also for unlocking new possibilities in diverse fields. Microresonator-based optical sensors[1,2] are excellent examples in this sense, which have been widely used in the detection of various biological and chemical substances in which very small changes in the surrounding medium cause the resonance wavelength to shift, allowing high-resolution detection. The prospect of transforming optical sensors into wearable devices[22] holds the key to revolutionizing healthcare, diagnostics, and personalized monitoring; however, achieving this requires advanced technologies that enhance the performance of the sensors to detect minute changes in weak optical signals in real time. Existing measurement methods, such as optical spectrum analyzers or tunable diodes, often lack the necessary sensitivity, speed, and resolution in addition to being bulky. Alternatively, ultra-narrow linewidth tunable lasers are utilized in these measurements; however, they face challenges, including long-term (millisecond) frequency stability during the measurement duration and, in some cases, requiring dithering for stabilization[23,24]. Moreover, in addition to their high cost, their frequency is subject to some degree of uncertainty during a sweep. To overcome these problems, different approaches have been developed such as comparing the resonance width with sidebands of a modulated laser[25], or with a spectrum of an asymmetric fiber Mach-Zehnder interferometer (MZI)[26]; nevertheless, issues with stability and spectral resolution still persist. Moreover, in most of these studies, experimental findings were validated using lifetime measurement methods, particularly cavity ring-down techniques[27]. Another measurement technique combines a phase modulator with a vector network analyzer at the expense of increased system complexity and the need for extra components such as high-speed phase modulators, filters, and fiber amplifiers[28].

Addressing these limitations, we introduce a novel PIC measurement method termed the "Sweeping Optical Frequency Mixing Method (SOHO)"[29], which leverages the principles of optical frequency mixing[30] by incorporating a sweeping laser, a fixed-wavelength laser and custom control software. Using this approach, we measured the spectrum of an ~85m-long unbalanced MZI and demonstrated 19.1 fm of spectral resolution over 12 GHz bandwidth, with the added benefit of real-time operation. Finally, through characterization of ultrahigh-quality factor ($Q=46\times10^6$) optical resonators, we verified the superior performance of SOHO both in resolution and sensitivity compared to conventional methods based on a tunable laser and a photodiode. Finally, we compared its performance with a standard method using a state-of-the-art tunable laser system and demonstrated superior performance in precision, speed, and sensitivity. This new method not only holds great promise for PIC characterization but also for high-resolution spectral measurements of weak optical signals[31], including applications such as long-range telecommunications, wearable sensors, spectroscopy (e.g. Raman and Brillouin), and biomedical imaging.

## RESULTS

**Operation principle and measurement setup**
SOHO is a measurement method based on the optical frequency mixing process. As depicted in Fig. 1a, this method involves mixing the output of a sweeping laser (called "sample laser"), following the photonic device under test, with a reference laser on a detector. The resulting difference in optical frequencies appears as a distinct peak in the electric spectrum analyzer (ESA). This enables us to map the optical frequency of the sample laser relative to the reference laser into the radio frequency (RF) spectral domain, which can be accurately measured using electronics. The working principle of SOHO is illustrated in Fig. 1b, showcasing the measurement of an optical resonator where the envelope is superimposed on both the optical and RF signals. For the reference laser, we chose a non-scanning laser, i.e. fixed-wavelength laser, since it can remain stable and predictable during measurements compared to a scanning one. The frequency of the sample laser is compared with the frequency of the reference laser in the ESA, eliminating the need to determine the exact frequency of the sample laser. Moreover, the frequency deviations and noise of the sample laser can be automatically identified on the RF spectrum. This approach also allows for the utilization of a laser with a stable amplitude and random frequency fluctuations as the sample laser as it ensures robust and accurate measurements even in the presence of unpredictable frequency variations in the sample laser.

The experimental setup of SOHO is shown in Fig. 1a. The sample laser (EMCORE TTX1995 micro-ITLA) is coupled to the photonic microchip via a single-mode optical fiber after passing through a polarization controller. The light coming out of the microchip is sent through a 90/10 splitter and 90% of the light is sent to an amplified photodiode (Thorlabs, DET08CFC/M). This part of the setup is the commonly-used tunable laser-based characterization setup (termed "standard method") and we embedded it into the SOHO setup to be able to make a direct comparison between the two methods. The remaining 10% of the light is combined with the reference laser (Thorlabs, WDM8-C-23A-20nm) through a 50/50 coupler and is sent into a high-speed balanced detector (Thorlabs, BDX3BA). A polarization controller was employed after the reference laser. When the frequency difference of the sample and the reference lasers are within the bandwidth of the photodetector, a beat frequency equal to the difference between the frequencies of the two lasers is detected on the balanced photodetector and measured with the electrical spectrum analyzer (ESA, Signal Hound, BB60D). The output signal has an intensity proportional to the product of the amplitudes of the two lasers. SOHO allows for measuring a spectral window width that is twice the electrical bandwidth without the need to scan the reference laser. Control of lasers, reading the ESA output, and data processing are done using a custom code written in C#.

We proceeded to conduct a thorough characterization of SOHO, deriving its linearity performance parameters. To check the linearity of the signal amplitude as a function of input power, a variable optical attenuator (VOA, shown in Fig.1a), was used as the sample. Figure 1c depicts the ESA output (converted to linear scale using Eq. (4)) relative to the input power measured after the sample with a photodiode and calibrated with a power meter. As can be seen, it exhibits a good linear relationship ($R^2$=1) between input power and the measured signal. Although the short-term line width of the sweeping laser is less than 100 kHz, the presence of frequency dither causes its long-term (millisecond) linewidth to broaden to approximately 100 MHz (Fig. 1d), resulting in a flat-top spectral shape and limiting the spectral resolution of the standard method to around 0.8 pm. On the other hand, the electronics part of the measurement system comprising the photodiode, transmission cables, and RF amplifier of the ESA doesn't have a flat RF response. Figure 1e illustrates the system's RF response, determined by substituting the sample with an optical attenuator during the measurement, which is used to compensate for all measurements.

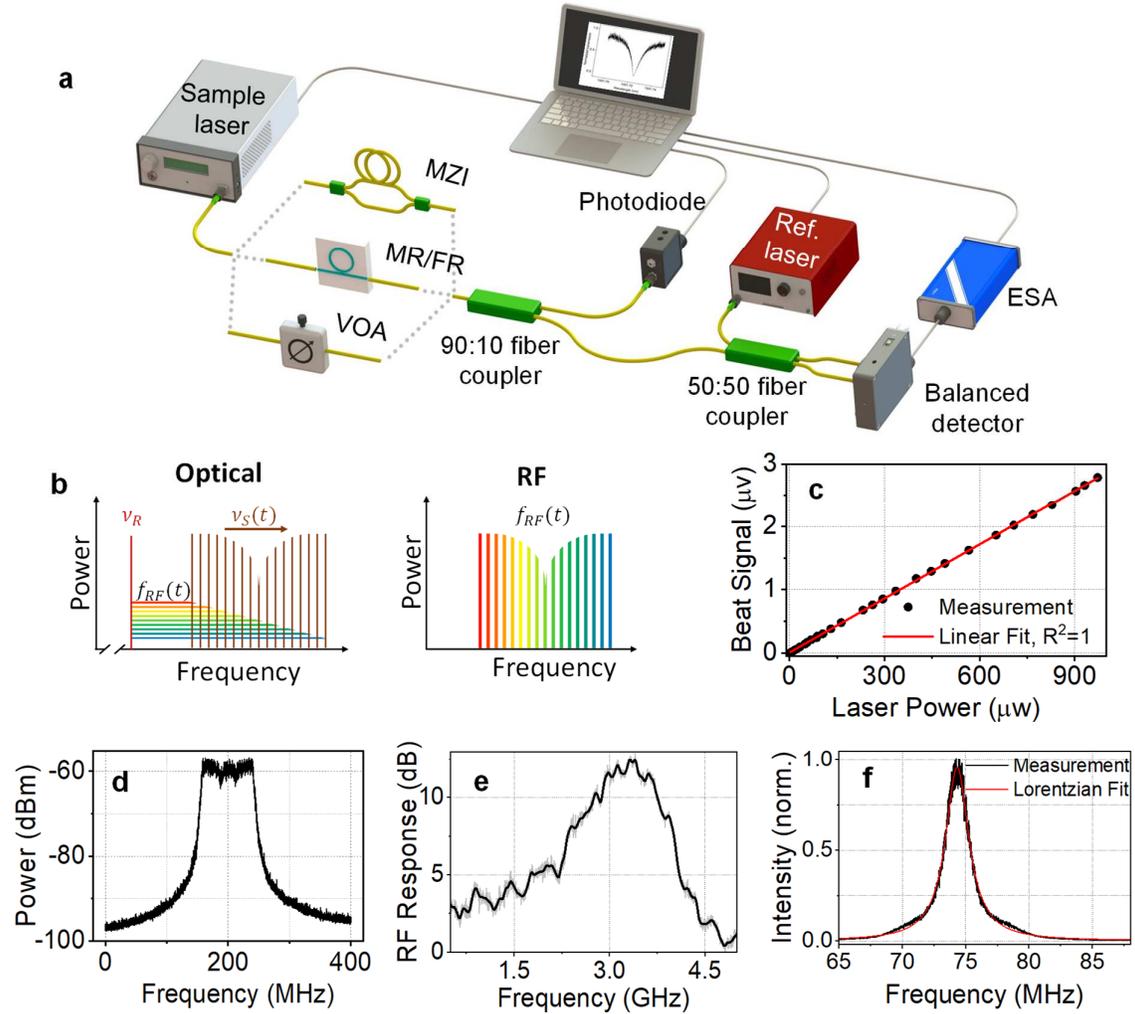

**Fig. 1. a.** Schematic of the experimental setup of SOHO. The (sweeping) sample laser (optical frequency $v_S(t)$) is coupled to the device under test and its output is combined with the reference laser (optical frequency $v_R$) via a 50:50 fiber coupler. The resulting interference beat-note is converted to an RF signal at $f_{RF} = |v_S(t) − v_R|$ and measured with the electric spectrum analyzer (ESA). The standard method (i.e. tunable laser with photodiode) is incorporated in this setup to make a direct comparison between the two methods. A variable optical attenuator (VOA) as a sample is used for checking the linearity of the measured signal relative to the input power. The output power of the sample port is calibrated using a power meter (not shown). An unbalanced fiber Mach-Zehnder interferometer (MZI) as a sample is used to show the resolution enhancement of SOHO. The response of microresonator/fiber-ring resonators (MR/FR) was also measured using the same setup. The polarization controllers that are placed after sample and reference lasers are not shown in this figure. **b.** depicts the frequency mixing process in the frequency domain for both input optical and output RF tones. The envelope on the optical and RF spectrum demonstrates the typical response of an optical resonator. **c.** The linearized beat signal on ESA relative to the sample laser power. The red dashed line is the linear fit of the data. **d**. Long-term linewidth measurement of the sample laser. **e.** RF response of the measurement system in dB scale, measured using VOA as the sample. Such a frequency response is used in all measurements for compensation. **f.**

The linewidth of the reference laser is measured with a frequency comb generated by a femtosecond laser as 2.2 MHz. The Lorentzian fit is given in red.

**Spectral measurements of racetrack resonators**

The microcavity resonators (racetrack) were fabricated using a hybrid waveguide platform that we recently developed[32] (see Fig. 2a, fabrication details are given in the Materials and Methods section). A comparison of the spectral measurement results of the fabricated racetrack resonators measured with the standard method and SOHO is given in Fig. 2e. The full-width-at-half-maximum (FWHM) of the resonance peak centered at 1550 nm was measured as 9.5 pm, which corresponds to an intrinsic quality factor of $Q_{int} = 2.7 \times 10^5$. The effective radius of the racetrack resonator was calculated as $R_{eff}$ = 166 µm using the formula given in Ref. [33]. By inserting $R_{eff}$ and $Q_{int}$ into Eq. (4) in Ref. [32], the optical loss value in the racetrack resonator was calculated as 1.3 dB/cm, which is mainly dominated by the absorption loss of the SU8 layer. To measure the transmission spectrum of the racetrack resonator using the SOHO setup, the reference laser is tuned to a frequency that is 4 GHz lower than the racetrack resonance frequency. The optical frequency of the sample laser is scanned between that of the reference laser and 6 GHz (ESA bandwidth) away from it in order to measure the transmission spectrum of the sample. Despite having a 5 GHz bandwidth, the balanced detector has sufficient responsivity to be used for 6 GHz measurements. To capture data while the sample laser is scanning, the ESA center frequency is put on the beat frequency and is moved with it during the scan, and data is saved with maximum holding. The span frequency of the ESA is set at 150 MHz, so the full width of the beat signal is always seen on the ESA measurement window.

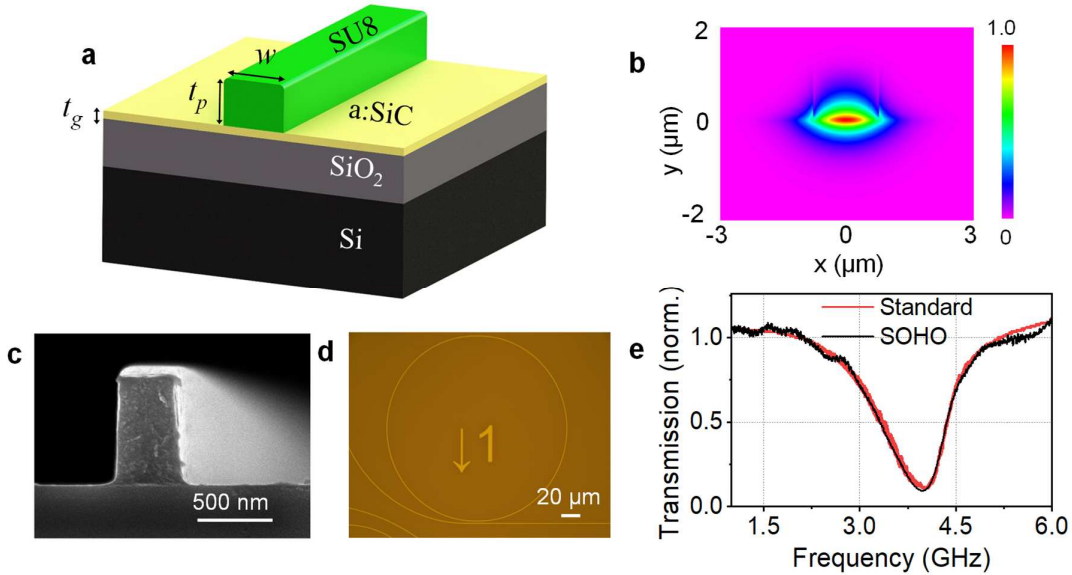

**Fig. 2. a.** Schematic of the hybrid waveguide structure, **b**. Mode profile of the fabricated waveguide. **c.** The scanning electron microscope image of the waveguide cross-section. **d.** Optical microscope image of the fabricated racetrack resonator. **e.** The transmission spectrum of the racetrack resonator that is obtained by using the standard method (red) and SOHO (black). The performance of SOHO is comparable with the standard method for the microresonators with a $Q < 10^6$.

**Spectral measurements of fiber ring resonators and Mach Zehnder interferometers**

In order to characterize the SOHO setup, we fabricated fiber ring resonators with different $Q$ values ($Q= 3\times10^6$ and $Q= 46\times10^6$) and an unbalanced MZI structure with a fiber length difference of $\Delta l =$ ~85m, corresponding to an FSR of 2.4 MHz in frequency or 19.1 fm in wavelength. The fiber ring resonators were fabricated by splicing one input of a 90/10 2×2 coupler with one of its outputs. The spectrum of a fiber ring resonator with a resonance full-width-at-half-maximum of around FWHM = 80 MHz (corresponding to a $Q = 3\times10^6$), a little less than the laser's long-term linewidth, was measured by the standard method and SOHO (Fig. 3a). Since SOHO provides higher resolution and dynamic range, it shows a sharper and deeper resonance dip while it washes out with the low resolution of the standard method, hence our new method is proven to be a more accurate spectral measurement method compared to the standard method.

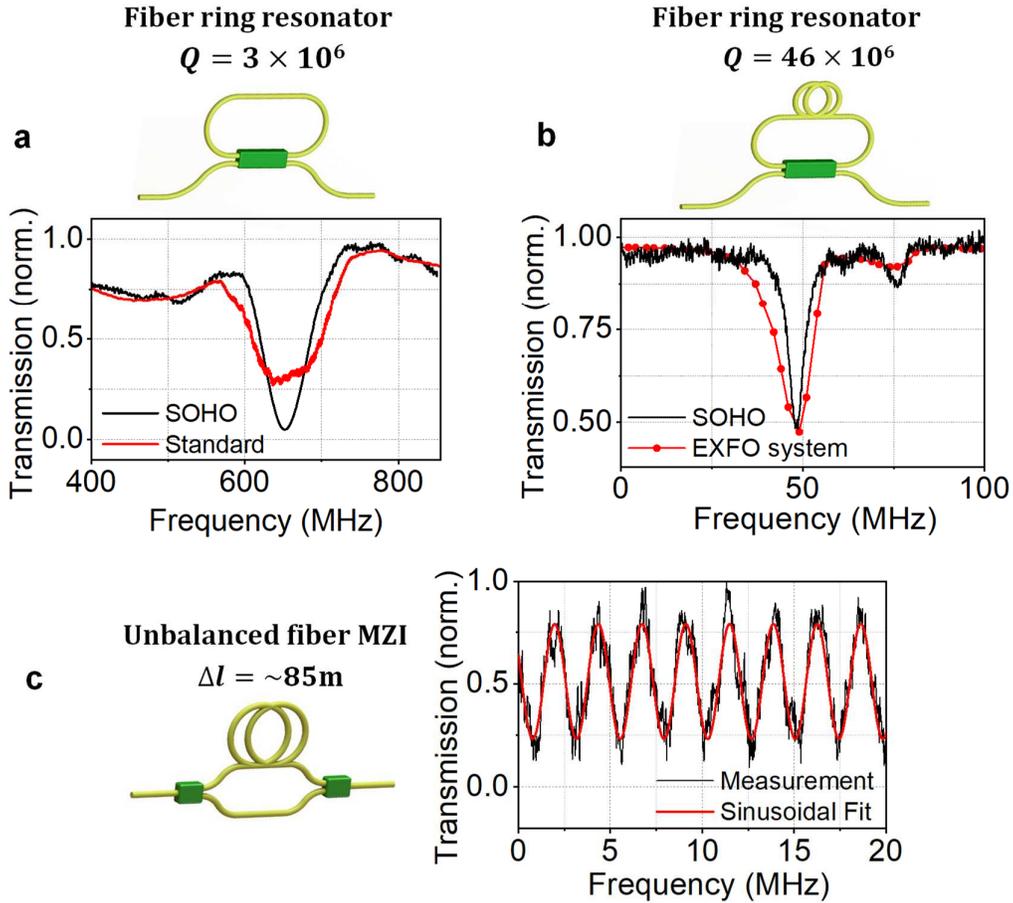

**Fig. 3.** Transmission spectrum of the fiber ring resonators with **a** $Q = 3\times10^6$ and **b** $Q = 46\times10^6$. In **a,** the measurements were done with SOHO (black) and the standard method (red) using the same sweeping laser in both cases. Because of the higher resolution and dynamic range of SOHO, it shows a better demonstration of the resonance spectrum compared to the standard method. In **b** the results of SOHO (red) are compared with the standard method using a state-of-the-art tunable laser system from EXFO (red). SOHO measures the spectrum with higher precision and accuracy. **c** Transmission spectrum of the fiber-based unbalanced MZI with $\Delta l = $ 85m, corresponding to an FSR= 19.1 fm. Extra noises on MZI data stem from capturing vibrations and acoustic noises by long fiber length during the measurement.

To further demonstrate the resolution enhancement, the transmission spectrum of a fiber ring resonator with a $Q= 46\times10^6$ was measured using SOHO and with the standard method where the sample laser was substituted with a state-of-the-art tunable laser system (EXFO, T500S-CLU+CTP10) and alterations were made to the detectors and compensation algorithms. To remove the polarization difference between the two setups and make measurements comparable, the device under test was placed between two polarizers in tandem with polarization controllers (not shown in Fig. 1). The measurement results, presented in Fig. 3b, reveal significant differences. The FWHM of the resonance peak was measured as 4.2 MHz using SOHO, whereas it becomes 9 MHz when the EXFO system is used. Additionally, capturing the spectrum depicted in Fig. 3b takes ~1 second with the EXFO system, while SOHO operates in real-time (See Supplementary Video 1, frame rate= 30 frames/s). These findings not only underscore the superior performance of our new method in terms of resolution, speed, and precision but also highlight its advantage over a cutting-edge tunable laser system.

The spectral measurements of the MZI device were performed by using both the SOHO setup and the standard method. SOHO clearly resolved the interference fringes, as depicted in Fig. 3c, while the standard method yielded a noise-like response (not shown in Fig. 3c). The linewidth broadening of the sample laser as demonstrated in Fig. 1d was the reason for the standard method's failure to resolve the interference fringes.

**Real-time operation of SOHO**

We further advanced the SOHO approach to be able to do real-time spectral measurements. Toward this goal, we dithered the output wavelength of the sample laser by triangular and sinusoidal signals using a function generator. In this configuration, the amplitude of the applied signal dictates the ultimate linewidth, while its frequency determines the dithering frequency. As the dwell time of the laser wavelength is higher near the minimum and maximum of the sinusoidal modulation signal, the spectrum has an M shape, whereas it becomes flatter for a triangular modulation signal as shown in Fig. 4a; however, it does not become completely flat. To correct for this problem and also to normalize the data by the RF response of the system (Fig. 1e), the measured spectrum of the photonic microchip is divided by a reference spectrum acquired by replacing the microchip with a VOA. Correcting nonlinear wavelength changes over time is challenging with the standard technique, especially for tunable diodes with mechanical sweeping mechanisms, as nonlinearity occurs during motor acceleration at the start and end of the sweeping range, complicating precise correction. Furthermore, the standard method exhibits nonlinearity at higher modulation frequencies, necessitating correction, while SOHO inherently corrects this issue, proving to be a significantly superior choice for real-time measurements.

In the experiments, sinusoidal and triangular waves with 200 kHz frequency and 50% symmetry were used. Real-time spectral measurement speed is highly dependent on the measurement's frequency span and with 6 GHz bandwidth ($\approx$48 pm) it is performed at a frame rate of around 4.15 frames/s (see Supplementary Video 2), which is limited by the capture rate of the ESA.

To showcase the real-time spectral measurement capabilities, we monitored the racetrack resonator's resonance shift as its temperature gradually increased by illuminating it from above with a 150 W lamp. Figure 4b presents the measurement setup and Figure 4c demonstrates eight snapshot images extracted from the real-time spectral measurement video, covering a 4-second interval while the microchip was being illuminated. These images visually demonstrate that as the microchip's temperature rises, its resonance consistently shifts towards lower beat frequencies, indicating a transition to shorter wavelengths.

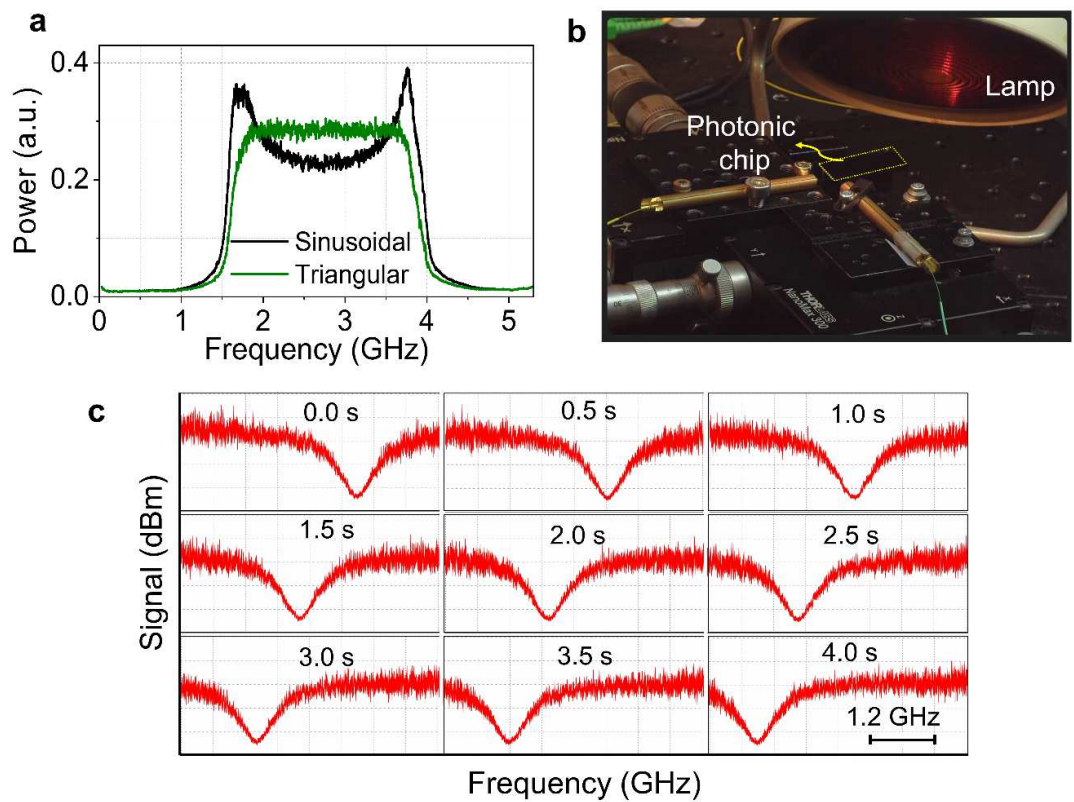

**Fig. 4**. **a.** Comparison of measured spectral shapes of the laser for the sinusoidal (black) and triangular (green) dithering signals. Here both of the plots are normalized to the RF frequency response of the system (Fig. 1e), which is the maximum holding of the ESA measurement data when the laser wavelength is swept slowly (10 mHz). **b.** The measurement setup used for the real-time measurements. The photonic chip comprising the racetrack resonator was illuminated by the lamp from above for 4 seconds. **c.** Snapshot images of the real-time video of the racetrack resonance shift during 4 seconds of lamp illumination. The video is provided in Supplementary Video 2.

## DISCUSSION

Since the SOHO approach relies on down-converted frequency detection on ESA, the precision and stability of the sample laser are inconsequential, as they will be referenced against the frequency of a stable laser. Furthermore, for the same reason, the linewidth of the sample laser does not have any effect on the frequency resolution of the measurement. In the current configuration of the experimental setup given in Fig. 1a, we deliberately used a tunable laser as the sample laser to be able to have a fair comparison between SOHO and the standard method; however, tunable laser is not compulsory for SOHO to deliver the same performance. Even employing a laser with a wider linewidth is more preferential as such a laser can allow for a faster full scan of the measuring frequency window.

Another advantage that stems from the innate amplification property of optical frequency mixing is that it can measure very weak signals with a good signal-to-noise ratio. This is important in particular for PIC devices with low fiber-chip coupling efficiency.

Additionally, measurements of nonlinear high-Q microcavity resonators can benefit from this approach dramatically as a few microwatts of optical power can excite unwanted thermal and nonlinear effects[34,35].

SOHO has potential particularly in optical sensing applications. For accurate measurement and continuous monitoring of resonance shifts in on-chip resonators, it is advantageous to employ a wide linewidth laser as a sample laser, aligning its wavelength with the resonator's resonance. This configuration makes it possible to view the resonance spectrum without scanning the laser frequency and thereby it substantially enhances the measurement speed, limited only by the maximum scan rate of the ESA. Hence, with this method, we can employ a non-scanning, stable laser as a reference along with a cost-effective, free-running sweeping laser that doesn't necessitate frequency stabilization, providing a practical advantage for wearable devices. Furthermore, the innate amplification and high sensitivity of our method relaxes the stringent requirements for input and output coupling, enabling the development of disposable optical biosensors. Figure 5a shows the measurement configuration that was used to characterize a photonic microchip comprised of several racetrack resonators. Despite a significant gap of approximately 1 mm between the photonic microchip and the input/output fibers, we were able to measure the spectrum with ~20 dB signal-to-noise ratio, thus validating the relaxed coupling requirements facilitated by our method (Fig. 5b). The distance between adjacent waveguides is 250 $\mu m$ while the beam radius of the fiber mode on the chip facet for this distance is less than 150 $\mu m$, therefore the crosstalk between adjacent waveguides is low. Considering that alignment issues represent a significant bottleneck in the advancement and widespread adoption of disposable optical sensors, our approach offers a distinct solution to address this challenge.

The theoretical spectral resolution of SOHO is determined as 6.2 fm by dividing the linewidth of the reference laser (2.2 MHz) by $2\sqrt{2}$, assuming a Lorentzian line shape and applying the Sparrow criterion. We experimentally demonstrated 19.1 fm, using the unbalanced fiber MZI. However, the long fiber arm's susceptibility to mechanical vibrations and thermal fluctuations hindered the demonstration of the theoretical spectral resolution. Such additional noise leads to very fast changes in the spectrum and distorts the real spectral shape.

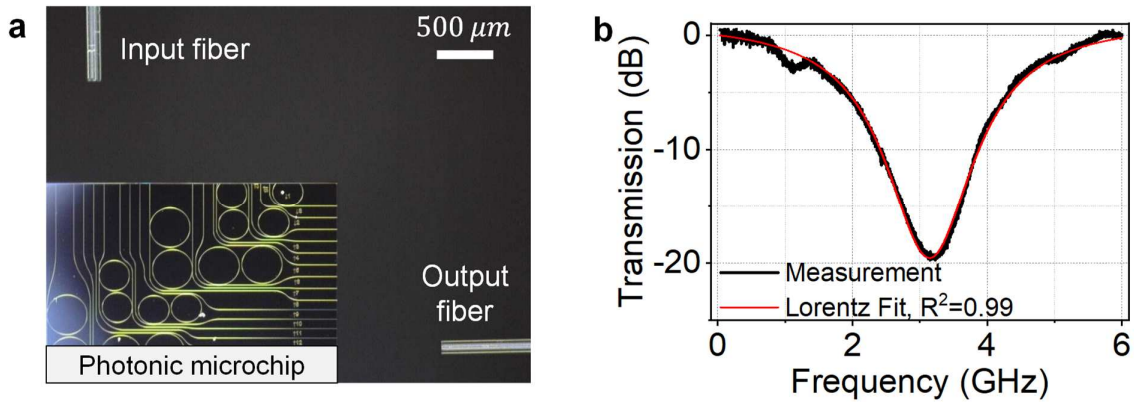

**Fig. 5**. **a.** The measurement configuration that was used to test the relaxed coupling requirements of SOHO. **b.** The spectrum of the racetrack resonator measured with the configuration given in **a**. Here the signal is ~20 dB higher than the noise floor.

In conclusion, SOHO offers the advantage of high sensitivity, speed, and precision in PIC measurements, making it a valuable tool for researchers and engineers working on the development, testing, and optimization of photonic devices and integrated circuits. Its excellent performance metrics hold significant promise for advancing current applications and unlocking new possibilities, including next-generation wearable devices, disposable point-of-care sensors, and many more.

**MATERIALS AND METHODS**

**Optical frequency mixing process:** Optical frequency mixing involves two optical signals, whereas the mixing product is an electrical signal. If the device under test is put after the sample laser, by sweeping the frequency of the sample laser, its optical transmittance spectrum, $T(\nu)$, can be extracted from the measured RF spectrum, $P_{log}(f)$. In this configuration the optical intensity after the interference of two lasers is:

$$I(\nu) \propto I_R + I_S + 2\sqrt{I_R I_S T(\nu)} \qquad (1)$$

where, $I_R$ and $I_S$ are the intensity of reference and sample lasers, respectively, and $T(\nu)$ is the transmittance of the sample as a function of optical frequency, $\nu$. The detected signal on the photodiode (proportional to the AC part of the intensity) is:

$$V(f) = \alpha\sqrt{I_R I_S T(\nu_R \pm f)} \qquad (2)$$

where $\alpha$ is a relative coefficient representing the optical to electrical conversion responsivity of the setup and $f$ is the electrical frequency derived from the optical down-conversion of the beat frequency between the lasers, i.e. $f = |\nu_S - \nu_R|$. The ESA measures electrical power spectral density as a function of $f$ and demonstrates it in the dBm scale. The following formulas represent it in linear scale:

$$P_{lin}(f) = \frac{V(f)^2}{Z} = \frac{\alpha^2 I_R I_S T(\nu_R \pm f)}{Z} \qquad (3)$$

$$P_{lin}(f) = \frac{10^{\frac{P_{log}(f)}{10}}}{10^3} \qquad (4)$$

where $Z$ is the input impedance of the ESA and varies based on the measurement conditions; i.e. $\nu_S > \nu_R$ or $\nu_S < \nu_R$, either the positive or negative sign is applied. By combining these two equations, $T(\nu)$ is obtained as:

$$T(\nu) = \frac{Z}{10^3 \alpha^2 I_R I_S} 10^{\frac{P_{log}(f)}{10}} \qquad (5)$$

**Microcavity resonator design and fabrication:** A 100-nm thick a-SiC layer was deposited on an 8-µm-thick thermally-oxidized silicon wafer using the inductively coupled plasma chemical vapor deposition (ICPCVD) method with an optimized recipe[36]. The schematic of the hybrid waveguide structure is given in Fig. 2a. The propagation loss of the a:SiC racetrack resonator based on this recipe was reported as 0.73 dB/cm. We measured the material absorption of the SU8 layer as ~3 dB/cm at 1550 nm wavelength. The refractive index of the thermal oxide, polymer layer, and a-SiC layer are 1.45, 1.58, and 2.57 at λ=1550 nm, respectively. The width was chosen as *w*=1.5 µm to guide only the fundamental mode. The thickness of the SU8 layer was 850 nm and an air cladding was used to reduce the loss induced by the SU8 layer. The thickness values of waveguide layers are calculated by using the waveguide design approach that we introduced in Ref. [32]. The devices were fabricated using e-beam writing. We did not use any conductive polymers to define the structures in contrast to common practice and we directly exposed the SU8 layer, which reduced the fabrication steps and eliminated the contamination of residuals of conductive polymer. To do so, we reduced the exposure dose of the e-beam writing (3uC/cm$^2$), and increased the writing speed (200 mm/s) as well as the resolution (10 nm). This step was followed by developing exposed samples with 1-methoxy-2-propanol acetate. We post-baked the fabricated devices at 170 °C for an hour using a hot plate in order to decrease the scattering losses. The measured width of the fabricated waveguides was 1.50±0.05µm. Fabricated devices were cleaved but facets were not polished. The mode profile of the fabricated waveguide is given in Fig. 2b. The optical microscope image of the fabricated racetrack resonators as well as the scanning electron microscope (SEM) image of the waveguide cross-section are given in Figs. 2d and 2c, respectively. The radius of the racetrack resonator was *R* = 150 µm and the length of the couplers was 50 µm.

## DATA AVAILABILITY

All raw and processed data that support the findings of this study are available from the corresponding author upon request.

## ACKNOWLEDGEMENT

This work was financially supported by the NWO Open Technology Program (COMBO, 18757).

## AUTHOR CONTRIBUTIONS

M.M.D built the SOHO system, designed and fabricated the fiber-loop resonators and MZI, performed the experiments, and wrote the manuscript. H.N. and M.T.K. fabricated the microresonators. N.S., B.L.R., and I.E.Z. deposited the high-quality SiC wafers B.I.A. designed the project, supported the processing of the experimental data, and revised the manuscript.

**Supplementary Information (videos)** is available for this paper.